\documentclass[11pt]{article}
\usepackage{amsmath,amsfonts}
\usepackage{mathtools}
\usepackage{multirow,multicol,array,bm}
\usepackage{tabularx, booktabs}
\usepackage{dsfont}
\usepackage[colorlinks=true,allcolors=blue]{hyperref}
\usepackage{tikz}
\usetikzlibrary{arrows.meta}
\usetikzlibrary{decorations.pathreplacing}
\usepackage{graphicx}
\usepackage[skip=0.333\baselineskip]{subcaption}
\usepackage[font=normal,labelfont=bf]{caption}
\usepackage[sort, numbers]{natbib}
\usepackage[a4paper, total={6in, 8in}]{geometry}
\usepackage{hyperref}
\newcommand{\footremember}[2]{%
\footnote{#2}
\newcounter{#1}
\setcounter{#1}{\value{footnote}}%
}

\begin{document}
\title{Estimating Potential Demand and Customers' Perception of Service Value in a Two-station Service System 
% \thanks{The authors would like to thank Opher Baron and Philipp Afech for their valuable comments and advice. This research was supported by the European Union’s Horizon 2020 research and innovation programme under the Marie Sklodowska-Curie grant agreement no.\ 945045, and by the NWO Gravitation project NETWORKS under grant agreement no.\ 024.002.003. \includegraphics[height=1em]{EULogo.jpg}}
}
% \titlerunning{Estimating in a Two-station Service System}

\author{
	Nishant Mangre \footremember{alley}{Korteweg-de Vries Institute for Mathematics, University of Amsterdam, Science Park
904, 1098 XH Amsterdam, The Netherlands. Email: nishant.mangre@student.uva.nl}%
	\and Jiesen Wang  \footremember{trailer}{Korteweg-de Vries Institute for Mathematics, University of Amsterdam, Science Park
904, 1098 XH Amsterdam, The Netherlands. Email: jiesenwang@gmail.com}
}
\maketitle              % 
\begin{abstract}
The potential demand in the market and customers' perception of service value are crucial factors in pricing strategies, resource allocation, and other operational decisions. However, this information is typically private and not readily accessible.
In this paper, we analyze a service system operating across two stations, each with its own customer flow. Customers arriving at the system are informed of the waiting times at both stations and can choose to either join the local station, switch to the other station, or balk. Our objective is to estimate the arrival rates at each station and customers' perceived service value based on the observed workloads at both stations.
A significant challenge arises from the inability to observe balking customers and the lack of distinction between local arrivals and customers switching from the other station, as the switching cost is unknown. To address this issue, we employ maximum likelihood estimation and validate the effectiveness of the estimator through a series of simulations.

% \keywords{Estimation \and Two-station queues \and Unobservable balking \and Switching cost}
\end{abstract}
\section{Introduction}
Hospitals, restaurants, and supermarkets often operate through multiple providers across various regions, aiming to meet the specific demands of each area. In today’s digital age, customers have greater access to real-time information, enabling them to make more informed decisions.
For instance, the UHNM Live waiting times mobile app\footnote{http://www.waittimes.uhnm.nhs.uk/} provides real-time waiting times for NHS A\&E and Walk-in centers, helping patients select the most convenient location. Similarly, numerous real-time waitlist apps for busy restaurants\footnote{https://www.tablesready.com/blog/waitlist-apps} are designed to reduce waiting times and improve service efficiency.
As a result, customers are increasingly aware of waiting times across different locations. They may choose to switch to another location if the payoff from switching outweighs the cost and inconvenience of waiting at their local one.

For a company manager, accessing the real-time workload at each station is relatively straightforward. However, customers' perceptions of service value and their associated switching costs are typically private and not directly observed. Consequently, the manager cannot determine whether a joining customer is local or has switched from another station. This poses a significant challenge in estimating the potential demand at each station. If few customers join, the manager cannot discern whether the demand is genuinely low or if many customers are choosing to balk.

The manager is interested in understanding customers' perceptions of service value and the potential demand, as this information is critical for addressing key decision-making challenges, such as:
\begin{itemize}
\item How many servers should be allocated to each station?
\item Is it profitable to raise the service fee, even if doing so necessitates increasing the service rate?
\end{itemize}

If low throughput is driven by low demand, shutting down some servers may be more cost-effective than lowering the service price to attract more customers. Conversely, if the potential demand at one station exceeds that of another, reallocating more servers to the higher-demand station may be the optimal strategy.

In this paper, we analyze a two-station queueing system where customers may not necessarily join the station they arrive at, or may choose not to join the system at all. As a result, the effective joining process observed by the manager differs significantly from the arrival processes at the individual stations. Our goal is to estimate customers' perception of service value and the potential demand at each station based on workload observations. To achieve this, we also need to estimate the switching cost. With estimates of the potential arrival rates at each station and customers' perceived reward value, the system manager can address key operational decisions, such as how to allocate servers at each station, what prices to set for different service rates, and whether it is necessary to upgrade or shut down certain servers.

The method employed in our work is maximum likelihood estimation, which involves deriving the likelihood function for three cases, depending on the workload difference between the two stations and the station to which a customer joins. It is important to note that an arriving customer may choose to balk, and a joining customer may have originally arrived at the other station.

This paper is organized as follows. 
Section~\ref{sec:literature} reviews the relevant literature from three perspectives. In section \ref{sec:model}, we provide a detailed description of the model and the observed data.
Then Section~\ref{sec:likelihood} examines three possible cases with the same observations and presents the likelihood function. Numerical results demonstrating the performance of our estimator are presented in Section~\ref{sec:numercial}.
Finally, Section~\ref{sec:conclusion} offers concluding remarks and outlines directions for future research, highlighting the challenges and methodological differences when considering alternative observation formats or model variations.

\section{Literature review} \label{sec:literature}

Our work is related to estimation methods based on customer behavior in queueing systems. The literature review is organized from three perspectives. First, we discuss how customers' private information can be obtained through various channels. Next, we introduce behavioral queueing models that analyze customer decision-making within queueing systems. Finally, we review estimation methods commonly used in queueing systems.

\subsection{Methods for estimating customer behavior}

Surveys are often used to collect data about customers' decisions. As pointed out by \cite{CY18}, surveys can be used to collect opinions and attitude data, that cannot easily be concluded from panel data. See \cite{A17} and \cite{MT17}, for how surveys are used in healthcare and traffic systems. Many studies have shown that waiting times and switch costs are significant variables for individuals’ willingness to join, and are used in explicit benefit expressions, see \cite{RMDO00} and \cite{D10} (figure 20).

However, surveys as a method for data collection also have many limitations, as has been pointed out by \cite{MC13}. It relies on answers to hypothetical questions, which may not accurately represent customer decision-making. 

Regression analysis can also determine customer perception of service value from panel data, for example, \cite{BUV21, D20, F23}. The benefit of regression-based estimation is that more variables can be considered since the calculation is more direct. This allows for comparing the significance of different explanatory variables, making it a powerful tool for explaining customers' decision-making. However, regression-based models sometimes assume that the potential arrival and switching cost are known (see \cite{G24}), and neglect the behavior of unobserved balking customers. 
%However, this information may be hidden and is important when deciding whether it is economically viable to upgrade the system.     

\subsection{Behavioral queueing models}
Many studies in operational research have considered behavioral queueing models, where individuals in the system can make decisions themselves. For a broad overview of different types of behavioral queueing models, see \cite{HH03, H16}. The first influential work on this topic is by \cite{N69}, where Naor studies customers' balking behavior in an M/M/1 queue, and concludes that social optima can be achieved by levying an admission toll. In his work, the cost of joining the queue is the expected waiting cost, which depends on the queue length and the service rate. Customers will balk if the expected waiting cost exceeds the reward of receiving the service. In \cite{PK04}, service durations are known at the time of customer arrival, and decision-makers are informed of the system workload, defined as the total remaining service time for all customers. That is, customers are aware of the exact waiting time when they make the decision. Our work assumes that the workload (waiting time for a customer if she joins) is observable, see \cite{H16} for more papers on strategic queueing with observable workload. In other cases, the available information is the queue lengths instead of the exact waiting time. See, for example, in \cite{HH94}, customers can purchase upon arrival the information on the queue lengths and decide which queue to join. 

L. Luski \cite{L76} was the first person to consider a model with multiple competitive queues. In his model, he considers a duopoly, in which two stations compete for customers. Each customer chooses a queue so that she minimizes her waiting time and service price. Many follow-up studies have researched this model, or similar models, for example, \cite{AF08, STY19}. These models are different from the model we consider since they assume that customers can switch freely between the different queues. In \cite{HG17}, customers join a queue and see its length. They can then choose to purchase the information on the other queue length and decide to switch or not.

\subsection{Estimation for queueing systems}
Even though many studies have analyzed queueing models, there are not many studies that have tackled estimation problems. As \cite{C58} has pointed out, these two fields are very different. In analytic studies, there is certainty about the properties of the model, whereas in statistical inference there exists an explicit uncertainty about the model and its parameters. A recent overview of inference studies on queues has been made by \cite{ANT21}. This includes estimation methods for queues with balking customers, for example, \cite{GH09, MZ13}. 

The queue inference engine is a method proposed by \cite{L90}. This estimation method relies only on transactional data. This means that only data about customers who choose the service is available. In \cite{DS93} and \cite{J99}, this method has been used to study queues with balking. However, nobody has yet used the queue inference engine in a setting with two or more competitive stations.

We will use a method similar to \cite{IRM23}, where the authors estimate the parameters of the customers’ patience-level distribution and the corresponding potential arrival rate, for an $M/G/s$ queue with unobserved balking. We consider a two-station queueing model with unobservable balking, and allows for switching between the stations at an immediate cost. The goal of this paper is to estimate the regional demands, customers' perception of service value, and the switching cost.

\section{The two-station queueing system} \label{sec:model}

We consider a system with two \textit{working stations}, labeled $s=1,2$. Each station contains one or more servers providing the same type of service. Potential customers arrive at station $s$ according to a Poisson process with rate $\lambda_s$. Each customer has a service value in mind, which is independently and identically distributed (i.i.d.). Let $H$ be the service value distribution, and $\tilde{H}$ be the tail function. Specifically, $\tilde{H}(x)$ gives the probability that a customer joins a station, if the cost of joining the system is $x$. We consider $H$  in a parametric framework and denote it as $H_{\theta}$, with $\tilde{H}$ written as $\tilde{H}_{\theta}$. Upon arriving at a station, customers observe the waiting times at both stations. The cost of joining a station consists of the waiting time plus a switching cost $c$, if the customer chooses a server other than the one she arrives at. Based on the perceived service value and the joining cost, each arriving customer decides to:
\begin{itemize} \item join the station where they arrive at,
\item switch to the other station, incurring a switching cost $c$, 
\item balk (decide not to join either station).
\end{itemize}
Service times depend only on the station a customer joins. The cumulative distribution function of the service time is  $G_1$ for Station 1, and $G_2$ for Station 2. Let $V_s(t)$ denote the waiting time if a customer joins station $s$ at time $t$. Customers will join the station with the lower total cost, defined as the sum of the waiting time and the switching cost (if applicable), and will balk if this cost exceeds their perceived service value. If the $j$-th customer arrives at station $s$ at time $t$, then she will
\begin{equation} \label{decision} \begin{dcases}
    \text{join station } s \ &\text{ if } V_s(t^-) \le V_{-s}(t^-) +c\  \wedge \  V_s(t^-) < R_{j},\\
    \text{join station } -s \ &\text{ if } V_s(t^-) > V_{-s}(t^-)+c\ \wedge\ V_{-s}(t^-) +c < R_{j},\\
    \text{balk} \ &\text{ if } \min(V_s(t^-),V_{-s}(t^-)+c) \geq R_{j},
\end{dcases} \end{equation}
where $-s\in\{1,2\}$ denotes the station other than $s$, $R_j$ is the service value for the $j$-th customer, and $\wedge$ denotes the logical AND.

Arriving customers who choose to join are referred to as \textit{effective arrivals}. Since some customers may choose to balk, the distribution of inter-arrival times for effective arrivals differs significantly from that of potential arrivals. Likewise, the distribution of customers who arrive at station $s$ and those who ultimately join station $s$ also differ, as customers may switch between the two stations

Let $k \in \mathbb{N}$ be the label of effective arrivals. We define $(A_k)_{k=1,2,\dots}$ as the sequence of inter-arrival time between the $k$-th joining customer and $k-1$-th joining customer, with $A_1$ being the arrival time of the first effective arrival. Let $(I_k)_{k=1,2,\dots}$ be the sequence of stations joined by the effective arrivals. Both $A_k$ and $I_k$ are important in our next stage analysis, and we need the following variables to determine $A_k$ and $I_k$. 

\begin{table}[h!]\caption{Table of Notation}
\begin{center}
\begin{tabular}{r c p{10cm} }
\toprule
$s$ & $:=$ & The label of each station, $s \in \left\{1,2\right\}$. \\
$-s$ & $:=$ & The label of the other station, other than $s$. If $s = 1$, then $-s = 2$; if $s = 2$, then $s = 1$.\\
$\lambda_{s}$ & $:=$ & Arrival rate at station $s$
 \\
$H_\theta$ & $:=$ & Cumulative distribution function, in terms of parameter $\theta$, of customers' perception of the service value. \\
$G_s$ & $:=$ & The service time distribution at station $s$. \\
$V_s(t)$ & $:=$ & The waiting time of a customer joining station $s$ at time $t$.\\
\multicolumn{3}{c}{}\\
\multicolumn{3}{c}{\underline{Parameters related to the arriving customers}}\\
\multicolumn{3}{c}{}\\
$j$ & $:=$ & index of arriving customers \\
${T}_{j}$ & $:=$ & Inter-arrival time between the $j$-th arrival and $j-1$-th arrival. \\
$S_j$ & $:=$ & The label of the station where the $j$-th customer arrives at. \\  
$R_j$ & $:=$ & The perceived service value of the $j$-th customer. \\
$B_j^{(s)}$ & $:=$ & The service time of the $j$-th customer if she chooses station $s$. \\
\multicolumn{3}{c}{}\\
\multicolumn{3}{c}{\underline{Parameters related to the effective arrivals}}\\
\multicolumn{3}{c}{}\\
$k$ & $:=$ & index of effective arrivals \\
$A_k$ & $:=$ & the inter-arrival times between the $k$-th effective arrival and $k-1$-th effective arrival. \\
$I_k$ & $:=$ & the station where the $k$-th effective arrival joins. \\
$X_k$ & $:=$ & the service time of the $k$-th effective arrival. \\
$H_\theta$, $\tilde{H}_\theta$ & $:=$ &the cumulative distribution function and tail function of customers' perception of the service value, in terms of parameter $\theta$. \\
\bottomrule
\end{tabular}
\end{center}
\label{tab:notation}
\end{table}

Let $j$ be the index of arriving customers, and $T_j$ be the inter-arrival time between the $j$-th arrival and $j - 1$-th arrival, with $T_1$ being the arrival time of the first customer. We use $S_j$ to denote the station the $j$-th customer arrives at. Note that an arriving customer may choose to balk or join the other station, which makes the difference between $S_j$ and $I_k$. Let $R_j$ be the perceived service value of the $j$-th customer, and $B_j^{(s)}$ be the service time of the $j$-th customer if she chooses station $s$. The notations used in the paper are represented in Table \ref{tab:notation}.

We assume that at $t=0$ both queues are empty, and $H_\theta(x) = 1$ if $x \leq 0$. This means that the first potential customer also becomes the first effective arrival, $A_1 = T_1$ and $I_1=S_1$. If we let $j_k$ be the index of $j$ corresponding to the $k$-th effective arrival. Then we know $j_1 = 1$, and for $k \geq 2$, 
\begin{equation}
j_k = \inf\left\{ j> j_{k-1}: \min\left\{V_{S_j}\left( \sum_{l=1}^j T_l^-\right),\ V_{-S_{j}}\left( \sum_{l=1}^j T_l^-\right)+c\right\} \le R_j \right\}.
\end{equation} 
Thus $A_k = \sum_{j =j_{k-1}+1}^{j_k} T_j$, and 
\begin{equation} 
I_k := \begin{dcases} S_{j_k}\quad &\text{ if } V_{S_{j_k}}\Big(\sum_{l=1}^{j_k} T_l^-\Big) \le V_{-S_{j_k}}\Big(\sum_{l=1}^{j_k} T_l^-\Big)+c, \\-S_{j_k} \quad &\text{ if } V_{S_{j_k}}\Big(\sum_{l=1}^{j_k} T_l^-\Big) > V_{-S_{j_k}}\Big(\sum_{l=1}^{j_k} T_l^-\Big)+c.
\end{dcases}
\end{equation}
The $k$-th effective arrival leads to an upward jump of size $X_k = B^{(I_k)}_{j_k}$ in the workload for station $I_k$.   

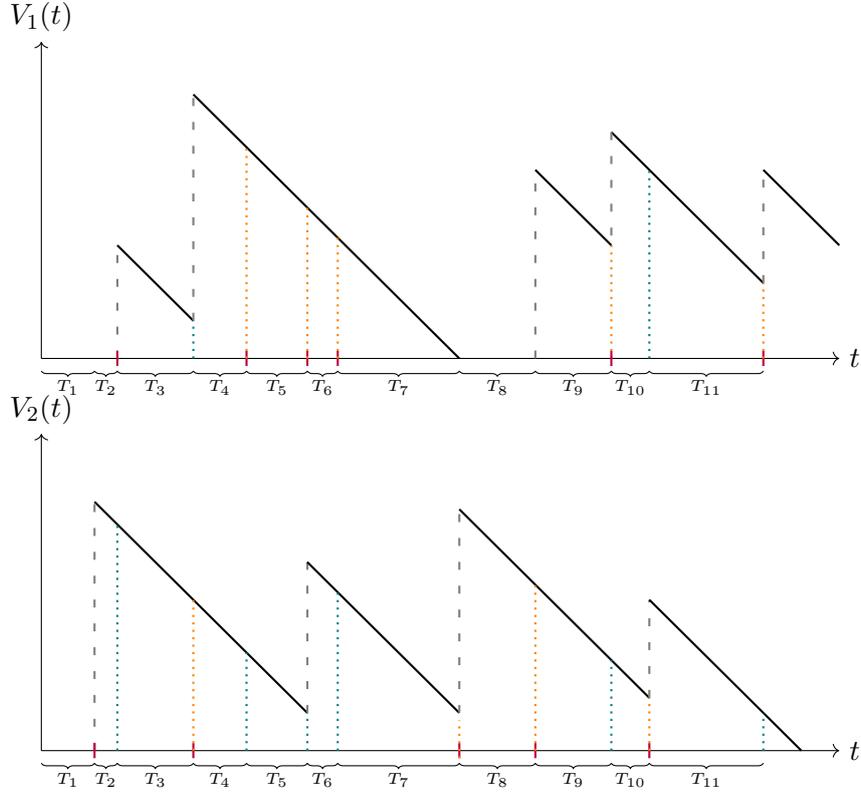
\begin{figure}[h]
    \centering
    \begin{tikzpicture}
        \draw[->] (-0.5, 0) -- (10, 0) node[right] {$t$};
  \draw[->] (-0.5, 0) -- (-0.5, 4.2) node[above] {$V_1(t)$};
  \draw[domain=0.5:1.5, smooth, variable=\x, thick] plot ({\x}, {-\x + 2});
  \draw[loosely dashed, thick, gray] (0.5, 0) -- (0.5, 1.5);
  \draw[domain=1.5:5, smooth, variable =\x, thick] plot ({\x}, {-\x + 5}); 
  \draw[loosely dashed, thick, gray] (1.5, 0.5) -- (1.5, -1.5  +5);
  \draw[thick, domain=6:7, smooth, variable=\x] plot ({\x}, {-\x + 8.5});
  \draw[loosely dashed, thick, gray] (6, 0) -- (6, 8.5 - 6);
  \draw[domain=7:9, thick, smooth, variable=\x] plot ({\x}, {-\x + 7 +3});
  \draw[loosely dashed, thick, gray] (7, -7 + 7 + 3) -- (7, 1.5);
  %\draw[dotted, thick, gray] (7, 1.5) -- (7, 0);
  \draw[domain=9:10, thick, smooth, variable=\x] plot ({\x}, {-\x + 9 + 2.5});
  %\draw[dotted, blue] (9, 0) -- (9, 2.5);
  \draw[loosely dashed, thick, gray] (9, -7 + 7 + 1) -- (9, 2.5);

  \draw[->] (-0.5, -5.2) -- (10, -5.2) node[right] {$t$};
  \draw[->] (-0.5, -5.2) -- (-0.5, 4.2 -5.2) node[above] {$V_2(t)$};
  \draw[domain=0.2:3, smooth, variable=\x, thick] plot ({\x}, {-\x -5.2 + 3.5});
   \draw[];
  \draw[loosely dashed, thick, gray] (0.2, -5.2) -- (0.2, -5.2 + 3.5-0.2);
  \draw[loosely dashed, thick, gray] (3, -3 -5.2 + 3.5) -- (3, -3 -5.2 + 4+ 1.5);
  \draw[domain=3:5, smooth, variable =\x, thick] plot ({\x}, {-\x + 5 - 5.2 + 0.5}); 
  \draw[loosely dashed, thick, gray] (5, -5.2 + 0.5) -- (5, -5.2 + 3.2);
  \draw[thick, domain=5:7.5, smooth, variable=\x] plot ({\x}, {-\x + 5 -5.2 + 3.2});
  \draw[loosely dashed, thick, gray] (7.5, -7.5 + 5 -5.2 + 3.2) -- (7.5, -5.2 + 2);
  %\draw[dotted, red] (8, -5.2) -- (8, -5.2 + 0.3);
  %\draw[dotted, red] (8, 0) -- (8, + 0.3);
  \draw[domain=7.5:9.5, thick, smooth, variable=\x] plot ({\x}, {-\x -5.2 + 1.5 + 8});

\draw[thick, purple] (0.2, -5.3) -- (0.2, -5.1); 
\draw[thick, purple] (0.5, -0.1) -- (0.5, 0.1) ;
\draw[thick, dotted, teal] (0.5, -5.2) -- (0.5, -5.2 + 3);

\draw[thick, purple] (1.5, -0.1-5.2) -- (1.5, 0.1-5.2) ;
\draw[dotted, thick, teal] (1.5, 0) -- (1.5, 0.5);
\draw[dotted, thick, orange] (1.5, -5.2) -- (1.5, -5.2 -1.5 + 3.5);

\draw[thick, purple] (2.2, -0.1) -- (2.2, 0.1);
\draw[dotted, thick, orange] (2.2, 0) -- (2.2, 5-2.2); 
\draw[dotted, thick, teal] (2.2, -5.2) -- (2.2, -2.2 -5.2 + 3.5);

\draw[thick, purple] (3, -0.1) -- (3, 0.1);
\draw[thick, dotted, teal] (3, -5.2) -- (3, -5.2 + 0.5);
\draw[thick, dotted, orange] (3, 0) -- (3, 2);

\draw[thick, purple] (5, -5.3) -- (5, -5.1);
\draw[thick, orange, dotted] (5, -5.3) -- (5, -5.3 + 0.5);

\draw[thick, purple] (6, -5.3) -- (6, -5.1);
\draw[thick, orange, dotted] (6, -5.2) -- (6, -6 + 5 -5.2 + 3.2);

\draw[thick, orange, dotted] (7, 0) -- (7, 1.5);
\draw[thick, purple] (7, -0.1) -- (7, 0.1);
\draw[dotted, thick, teal] (7, -5.2) -- (7, -7 + 5 -5.2 + 3.2);

\draw[thick, purple] (7.5, -5.3) -- (7.5, -5.1);
\draw[thick, orange, dotted] (7.5, -5.2) -- (7.5, -7.5 + 5 -5.2 + 3.2);
\draw[thick, teal, dotted] (7.5, 0) -- (7.5, 2.5);

\draw[thick, purple] (9, -0.1) -- (9, 0.1);
\draw[thick, dotted, orange] (9, 0) -- (9, 1);
\draw[thick, dotted, teal] (9, -5.2) -- (9, -9 -5.2 + 1.5 + 8);

\draw[thick, purple] (3.4, -0.1) -- (3.4, 0.1);
\draw[thick, orange, dotted] (3.4, 0) -- (3.4, -3.4 + 5);
\draw[thick, dotted, teal] (3.4, -5.2) -- (3.4, -3.4 + 5 - 5.2 + 0.5);

\foreach \i in {0, -5.2}{
  \draw [decorate,decoration={brace,amplitude=2pt,mirror,raise=1ex}]
  (-0.5,\i) -- (0.2,\i) node[midway,yshift=-1em]{\tiny $T_1$};
\draw [decorate,decoration={brace,amplitude=2pt,mirror,raise=1ex}]
  (0.2,\i) -- (0.5,\i) node[midway,yshift=-1em]{\tiny $T_2$};
\draw [decorate,decoration={brace,amplitude=2pt,mirror,raise=1ex}]
  (0.5,\i) -- (1.5,\i) node[midway,yshift=-1em]{\tiny $T_3$};
  \draw [decorate,decoration={brace,amplitude=2pt,mirror,raise=1ex}]
  (1.5,\i) -- (2.2,\i) node[midway,yshift=-1em]{\tiny $T_4$};
  \draw [decorate,decoration={brace,amplitude=2pt,mirror,raise=1ex}]
  (2.2,\i) -- (3,\i) node[midway,yshift=-1em]{\tiny $T_5$};
  \draw [decorate,decoration={brace,amplitude=2pt,mirror,raise=1ex}]
  (3,\i) -- (3.4,\i) node[midway,yshift=-1em]{\tiny $T_6$};
  \draw [decorate,decoration={brace,amplitude=2pt,mirror,raise=1ex}]
  (3.4,\i) -- (5,\i) node[midway,yshift=-1em]{\tiny $T_7$};
  \draw [decorate,decoration={brace,amplitude=2pt,mirror,raise=1ex}]
  (5,\i) -- (6,\i) node[midway,yshift=-1em]{\tiny $T_8$};
  \draw [decorate,decoration={brace,amplitude=2pt,mirror,raise=1ex}]
  (6,\i) -- (7,\i) node[midway,yshift=-1em]{\tiny $T_9$};
  \draw [decorate,decoration={brace,amplitude=2pt,mirror,raise=1ex}]
  (7,\i) -- (7.5,\i) node[midway,yshift=-1em]{\tiny $T_{10}$};
  \draw [decorate,decoration={brace,amplitude=2pt,mirror,raise=1ex}]
  (7.5,\i) -- (9,\i) node[midway,yshift=-1em]{\tiny $T_{11}$};
}
    \end{tikzpicture}
    \caption{An example of the queueing process, where the red ticks indicate the station each potential customer arrived at, and the dotted lines indicate the virtual waiting time they observed.}
    \label{fig:waitingtimeplot}
\end{figure}

Figure \ref{fig:waitingtimeplot} illustrates a queueing process. The red ticks on the horizontal line represent the arrival times. The orange and blue dotted lines indicate the virtual waiting times observed by the customer at their arrival station and the other station, respectively. In this example, 
\begin{itemize}
    \item The first customer arrives at and joins Station 2.
    \item The second customer arrives at and joins Station 1.
    \item The third customer arrives at Station 2 but finds that the waiting time at Station 1 plus the switching cost $c$ is less than the waiting time at Station 2. Therefore, she joins Station 1.
    \item The fourth customer arrives at Station 1 but chooses to balk.
\end{itemize}
Our objective is to estimate the arrival rate $\lambda_s$ and $\theta$, by observing the workload process $V_s(t)$ for $s\in\{1,2\}$. That is, in Figure \ref{fig:waitingtimeplot}, we only observe the black lines in each station, but not the red ticks. 

\begin{figure}[h!]
    \centering
    {\includegraphics[width=0.7\linewidth]{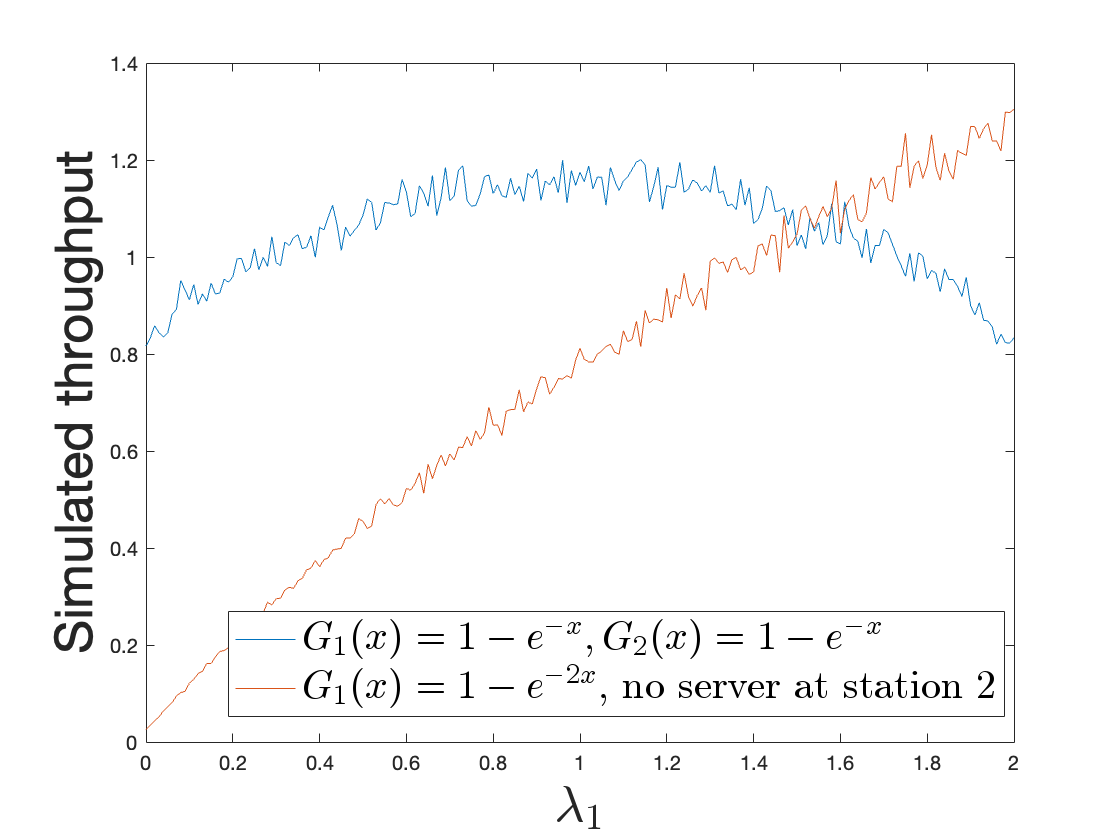}}
    \caption{The mean throughput per unit time for different values of $\lambda_1$, given that $\lambda_1+\lambda_2 = 2, \, c = 2, \, \tilde{H}_{\theta}(x) = \frac{1}{(x+1)^4}$.} \label{fig:example}
\end{figure}

We use a simulation example in Figure \ref{fig:example} to illustrate that estimation of $\lambda_s$ is important from the managerial perspective. In the example, $\lambda_1 + \lambda_2 = 2$. The system manager operates two servers, each with an exponentially distributed service time at rate 1. Arriving customers observe the lower joining cost between the two stations and choose to join with probability $(x+1)^{-4}$ when the lower cost is $x$. When $\lambda_1 > \lambda_2$, it may be advantageous to allocate both servers to one station to improve overall throughput. However, without knowledge of $\lambda_1$,  this decision cannot be validated. We simulated the process until 1000 effective arrivals were observed under two configurations: (I) one server at each station, (II) two servers at Station 1 with no server at Station 2. The mean throughput was calculated and plotted in Figure \ref{fig:example}. The blue line represents configuration (I), while the orange line represents configuration (II). It is evident that the orange line outperforms the blue line when $\lambda_1$ is larger than a value close to 1.5. This indicates that assigning both servers to Station 1 is only more effective than an even distribution when $\lambda_1$ is greater than a certain threshold.

\section{Likelihood function} \label{sec:likelihood}

The goal is to estimate the potential arrival rates $\lambda_1$ and $\lambda_2$, as well as the distribution parameter of the service value, $\theta$. To achieve this, we employ maximum likelihood estimation, which involves first expressing the likelihood function in terms of the unknown parameters.

For the system manager, it is reasonable to assume that she can observe the workload flow $V_s(t)$ at each station. The information contained in these workload flows includes $(A_k)$, $(I_k)$, and $(X_k)$. For example, in Figure \ref{fig:waitingtimeplot}, the jump times correspond to $A_k$, the station at which the jump occurs represents $I_k$, and the size of the jump indicates $X_k$.

Let $\tilde{A}_k:= \sum_{i=1}^k A_k$ be the time of joining for the $k$-th effective arrival. Assume that the observation contains the first $K$ effective arrivals, then the waiting time for the $k$-th effective arrival if she joins station $s$ is $V_s(\tilde{A}_k^-)$.

We will now consider the conditional likelihood of the observation $(A_k,\ I_k)$, given the observations $v^{(s)}_{k-1} := V_s(\tilde{A}_{k-1}^-) + X_{k-1} \mathds{1}_{\{I_{k-1} = s\}}$ for $s \in \{1,2\}$, with $v_0^{(1)} = v_0^{(2)} = 0$. We can write this density as 
\begin{align} \label{eq:pdf}
    &f_{A_k, I_k \mid v^{(1)}_{k-1}, v^{(2)}_{k-1}}(a,i) \\
    =  &\lim_{\Delta t \to 0}  \frac{1}{\Delta t} P(A_k \in [a, a + \Delta t), I_k = i \mid A_k > a,V_s(\tilde{A}_{k-1}^-) + X_{k-1} \mathds{1}_{\{I_{k-1} = s\}} = v^{(s)}_{k-1}) \notag\\ 
    &\cdot P(A_k \geq a \mid V_s(\tilde{A}_{k-1}^-) + X_{k-1} \mathds{1}_{\{I_{k-1} = s\}} = v^{(s)}_{k-1}) \notag \,.
\end{align} 

During $[\tilde{A}_{k-1}, \tilde{A}_{k-1} + a)$, each customer arriving at station $s$ chooses to join the system with probability $\max\{\tilde{H}(V_s(\tilde{A}_{k-1} + t)),\tilde{H}(V_{-s}(\tilde{A}_{k-1} + t)+c)\}$ for $0< t < A_k$. Thus customers become effective according to a time-inhomogeneous Poisson process with a time-dependent rate 
\begin{equation}\label{timerate}
    \begin{split}
        &\lambda_1 \max\{\tilde{H}(V_1(\tilde{A}_{k-1} + t)),\tilde{H}(V_{2}(\tilde{A}_{k-1} + t)+c)\}\\ + &\lambda_2 \max\{\tilde{H}(V_1(\tilde{A}_{k-1} + t)+c),\tilde{H}(V_{2}(\tilde{A}_{k-1} + t))\} .
    \end{split}
\end{equation}
To determine the exact rate, we will distinguish between three different cases: (i) $|v^{(i)}_{k-1} -v^{(-i)}_{k-1}| \le c$, (ii) $v^{(i)}_{k-1} > v^{(-i)}_{k-1} +c$, (iii) $v^{(i)}_{k-1} +c \leq v^{(-i)}_{k-1}$. Now we analyze the three cases one by one.
\begin{figure}
    \centering
    \subcaptionbox{$v^{(1)}_{k-1}-v^{(2)}_{k-1} < c$}{\includegraphics[width=0.45\linewidth]{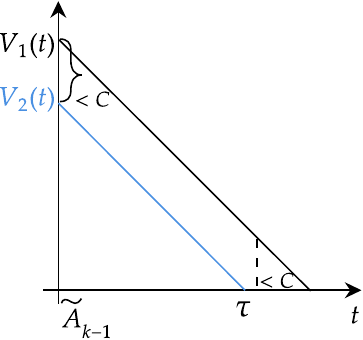}}
    \subcaptionbox{$v^{(1)}_{k-1}-v^{(2)}_{k-1} > c$}{\includegraphics[width=0.45\linewidth]{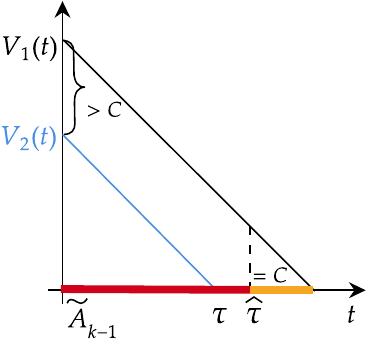}}
    \caption{Virtual waiting time against time}  \label{fig:TrvaelCostIllustrate}
\end{figure}

(i) When  $|v^{(i)}_{k-1} -v^{(-i)}_{k-1}| \le c$, then it must hold that \[|V_1(\tilde{A}_{k-1}+t) -V_2(\tilde{A}_{k-1}+t)| \le c, \qquad 0 < t < A_{k}.\] To see this, note that the virtual waiting time decreases linearly with time until it reaches zero or someone joins a queue. In Figure \ref{fig:TrvaelCostIllustrate}(a),  the workload difference is less than $c$ at $\tilde{A}_{k-1}$, and the workload at station 2 reaches $0$ at $\tau$, making the difference even less. In other words,
\begin{equation}\label{Vlinear}
    V_s(\tilde{A}_{k-1}+t) = \max(0,V_s(\tilde{A}_{k-1})-t),
\end{equation}
holds for all $0 <t <A_k$. We can use this formula to deduce that $$\tilde{H}(V_s(\tilde{A}_{k-1} + t)) = \tilde{H}(V_s (\tilde{A}_{k-1}) - t),$$ since $\tilde{H}(x) = \tilde{H}(0) = 1$, for all $x\le 0$. 
Using \eqref{timerate}, we find that the effective arrival rate at this time $t$ equals
\begin{equation} 
\lambda_1 \tilde{H}(v^{(1)}_{k-1} - t) + \lambda_2 \tilde{H}(v^{(2)}_{k-1} - t),
\end{equation}
and the probability 
\begin{align}
    & P(A_k \in [a, a + \Delta t), I_k = i \mid A_k > a,V_s(\tilde{A}_{k-1}^-) + X_{k-1} \mathds{1}_{\{I_{k-1} = s\}} = v^{(s)}_{k-1}) \qquad  0 <t <A_k \notag \\
    = &  \Delta t \, \lambda_i \, \tilde{H}(v^{(i)}_{k-1}-t) \,. 
\end{align}

(ii) When $v^{(i)}_{k-1} > v^{(-i)}_{k-1} +c$, the joining cost at station $-i$ is lower at time $\tilde{A}_{k-1}$. We find that if $t<A_k-c$, then $V_i(\tilde{A}_{k-1}+t) > V_{-i}(\tilde{A}_{k-1}+t) +c$. This means that the effective arrival rate during this time is
    \[
    \lambda_i\tilde{H}((v^{(-i)}_{k-1}-t)^+ +c ) +\lambda_{-i}\tilde{H}(v^{(-i)}_{k-1}-t ) \,,
    \]
and the probability of joining station $i$ is $0$. 

If $t\ge A_k-c$, then it holds that $|V_i(\tilde{A}_{k-1}+t) - V_{-i}(\tilde{A}_{k-1}+t) |\le c$, which means the arriving customers will choose the local station. Thus, the effective arrival rate during this time is
\begin{equation}   \lambda_1\tilde{H}(v^{(1)}_{k-1}-t ) +\lambda_2\tilde{H}(v^{(2)}_{k-1}-t ),
\end{equation} 
and the probability 
\begin{align}
    & P(A_k \in [a, a + \Delta t), I_k = i \mid A_k > a,V_s(\tilde{A}_{k-1}^-) + X_{k-1} \mathds{1}_{\{I_{k-1} = s\}} = v^{(s)}_{k-1}) \notag \\
    = &  \Delta t \, \lambda_i \, \tilde{H}(v^{(i)}_{k-1}-t) \,.
\end{align}
Figure \ref{fig:TrvaelCostIllustrate}(b) illustrates this situation if we treat $i$ as $1$. When $\tilde{A}_k > \tau$, the workload difference starts decreasing and reaches $c$ when $\tilde{A}_k = \hat{\tau}$. In the horizontal line, the red part represents $\tilde{A}_k < \hat{\tau}$, which is the case $t < A_k-c$, and the orange part represents $t \geq A_k-c$.

(iii) When $v^{(i)}_{k-1} + c < v^{(-i)}_{k-1}$, the joining cost at station $i$ is lower for arriving customers at both stations, at time $\tilde{A}_{k-1}$.
If $t<A_k-c$, then arriving customers all choose station $i$, so the effective arrival rate is 
    \begin{equation} 
    \lambda_i\tilde{H}(v^{(i)}_{k-1}-t ) +\lambda_{-i}\tilde{H}((v^{(i)}_{k-1}-t)^+ +c )\,,
    \end{equation}
and the probability 
\begin{align}
    & P(A_k \in [a, a + \Delta t), I_k = i \mid A_k > a,V_s(\tilde{A}_{k-1}^-) + X_{k-1} \mathds{1}_{\{I_{k-1} = s\}} = v^{(s)}_{k-1}) \notag \\
    = &  \Delta t \, \left(\lambda_i\tilde{H}(v^{(i)}_{k-1}-t) +\lambda_{-i}\tilde{H}((v^{(i)}_{k-1}-t)^++c)\right) \,.
\end{align}

If $t\ge A_k-c$, then the arriving customers will choose the local station, thus the effective arrival rate is
    \begin{equation} 
    \lambda_1\tilde{H}(v^{(1)}_{k-1}-t) +\lambda_2\tilde{H}(v^{(2)}_{k-1}-t) \,,
    \end{equation} 
and the probability 
\begin{align}
    & P(A_k \in [a, a + \Delta t), I_k = i \mid A_k > a,V_s(\tilde{A}_{k-1}^-) + X_{k-1} \mathds{1}_{\{I_{k-1} = s\}} = v^{(s)}_{k-1}) \notag \\
    = &  \Delta t \, \lambda_i \,  \tilde{H}(v^{(i)}_{k-1}-t)\,.
\end{align}
Figure \ref{fig:TrvaelCostIllustrate}(b) illustrates this situation if we treat $i$ as $2$. 

It follows from \eqref{eq:pdf} that
\begin{itemize}
    \item (i) when  $|v^{(1)}_{k-1} -v^{(2)}_{k-1}| \le c$,
    {\small
    \begin{equation}  
    f_{A_k, I_k\mid v_1,v_2}(a,i)\\
    = \lambda_{i}\tilde{H}(v^{(i)}_{k-1}-a) 
   \exp \bigg[-\lambda_{i} \int_0^{a} \tilde{H}(v^{(i)}_{k-1} - u) du -\lambda_{-i} \int_0^{a} \tilde{H}(v^{(-i)}_{k-1} - u) du\bigg];
    \end{equation}
    }
    
    \item (ii) when $v^{(i)}_{k-1} > v^{(-i)}_{k-1}+c$, 
    {\small
    \begin{equation}  \begin{split}
      &f_{A_k , I_k\mid v^{(1)}_{k-1}, v^{(2)}_{k-1}}(a,i) = \\[5pt]
    & \begin{dcases}
    0 & a < v^{(i)}_{k-1}-c,\\[8pt]
    \lambda_{i}\tilde{H}(v_{k-1}^{(i)}-a)\exp \bigg[
    -\lambda_{-i_k} \int_0^{a} \tilde{H}(v^{(-i)}_{k-1} - u) du & a \ge v^{(i)}_{k-1}-c; \\
    -\lambda_{i} \bigg( \int_0^{v^{(i)}_{k-1}-c} \tilde{H}((v^{(-i)}_{k-1}-u)^+ +c ) du  
    + \int_{v^{(i)}_{k-1}-c}^{a} \tilde{H}(v^{(i)}_{k-1} - u) du\bigg) \bigg] &
    \end{dcases}
\end{split} \end{equation} }

    \item (iii) when $v^{(i)}_{k-1}+c < v^{(-i)}_{k-1}$
    {\small
    \begin{equation} \begin{split} 
&f_{A_k, I_k \mid v^{(1)}_{k-1}, v^{(2)}_{k-1}}(a, i)\\[5pt]
    = & \begin{dcases}
    (\lambda_{i}\tilde{H}(v_{k-1}^{(i)}-a) +\lambda_{-i}\tilde{H}((v^{(i)}_{k-1}-a)^++c)) & a<v^{(-i)}_{k-1}-c,\\
     %\cdot \ 
     \exp \bigg[-\lambda_{i} \int_0^{a} \tilde{H}(v^{(i)}_{k-1} - u) du 
    -\lambda_{-i} \int_0^{a} \tilde{H}((v^{(i)}_{k-1}-u)^++c ) du \bigg] &\\[8pt]
    \lambda_{i}\tilde{H}(v^{(i)}_{k-1}-a) \exp \bigg[
    -\lambda_{i} \int_0^{a} \tilde{H}(v^{(i)}_{k-1} - u) du  &  a\ge v^{(-i)}_{k-1}-c.\\
    -\lambda_{-i} \bigg(\int_0^{v^{(-i)}_{k-1}-c} \tilde{H}((v^{(i)}_{k-1}-u)^+ +c ) du
    + \int_{v^{(-i)}_{k-1}-c}^{a} \tilde{H}(v^{(-i)}_{k-1} - u) du \bigg)\bigg] &
        \end{dcases}
\end{split} \end{equation} }
\end{itemize}

\begin{figure}[h!]
    \centering
    {\includegraphics[width=1\linewidth]{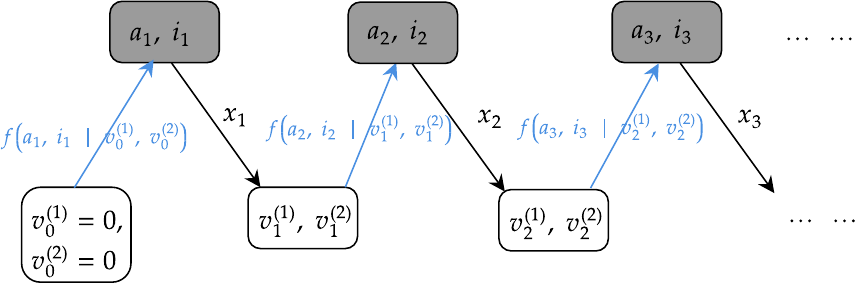}}
    \caption{Illustration of how the likelihood is derived.} \label{fig:flow}
\end{figure}

Let $\mathbf{A} = (a_1, \dots, a_k)$, $\mathbf{I}:= (i_1, \dots, i_k)$, and $\mathbf{X} = (x_1, \dots, x_k)$ be realizations for $A_1, \dots, A_k$, $I_1, \dots, I_k$, and $X_1, \dots, X_k$. We can now determine the full likelihood function 
\[
L_{K}(\lambda_1, \lambda_2, \theta, c; \mathbf{A, I, X}) \,.
\]
Note that the goal of deriving $L_K$ is to calculate the unknown parameters that maximize it, and the likelihood does not include $\mathbf{X}$, since $\mathbf{X}$ depends on the service time distribution, but not on balking or switching behavior. This means that the maximum likelihood estimator $(\hat{\lambda}_1, \hat{\lambda}_2, \hat{\theta}, \hat{c})$ for the observation, can be determined without considering the likelihood of $\mathbf{X}$, 
\begin{equation} \label{eq:MLE}
\begin{split}
(\hat{\lambda}_1, \hat{\lambda}_2, \hat{\theta}, \hat{c}) :=\ &\text{arg max }L_{K}(\lambda_1, \lambda_2, \theta, c; \mathbf{A, I, X})\\=\ &\text{arg max }L_{K}(\lambda_1, \lambda_2, \theta, c; \mathbf{A, I \mid X}).
\end{split}\end{equation}
Thus it is enough to have $L_{K}(\lambda_1, \lambda_2, \theta, c; \mathbf{A, I} \mid \mathbf{X})$. In Figure \ref{fig:flow}, we illustrate how the likelihood function can be derived. Given $v_0^{(1)} = v_0^{(2)} = 0$, we can compute $f(a_1, i_1 \mid v_0^{(1)}, v_0^{(2)})$. Then in addition with the value of $x_1$, we can infer the value of $v_1^{(1)}$ and $v_1^{(2)}$ and subsequently obtain $f(a_2, i_2 \mid v_1^{(1)}, v_1^{(2)})$, and so on. Thus we have
\begin{equation} 
L_{N}(\lambda_1, \lambda_2, \theta, c; \mathbf{A, I} \mid \mathbf{X}) = \prod_{n=1}^N f_{A_k, I_k\mid v_{k-1}^{(1)}, v_{k-1}^{(2)}}(a_k, i_k; \lambda_1, \lambda_2, \theta, c).
\end{equation}

\section{Numerical results} \label{sec:numercial}

In this section, we estimate $(\lambda_1, \lambda_2, \theta, c)$ using \eqref{eq:MLE}.
The estimation performance is demonstrated via simulations. Specifically, we use observations obtained from simulations, to estimate unknown parameters, and compare them with the parameters used in the simulations.

% We consider an example where the service value is exponentially distributed, that is $H_{\theta}(x) = 1- e^{-\theta x}$ for $\theta >0$. 
In the simulation, we tried exponential service time, where 
\[
G_s(x) = \mathbb{P}(X \leq x) =  1- e^{-\beta x}, \qquad \beta > 0, \quad \mbox{for} \quad x \geq 0
\]
and Pareto distribution, 
where 
\[
G_s(x) = \mathbb{P}(X \leq x) = 1- \frac{1}{(1+x)^\beta}, \qquad \beta > 0, \quad \mbox{for} \quad x \geq 0 \,.
\]
We assume that the service value is Pareto distributed,
that is $\tilde{H}_\theta = 1/(1+x)^\theta$.
We assume the initial state is empty, i.e. the virtual waiting times at both stations are $0$. In the simulation, we have 1000 runs, and in each run, we stopped the simulation after observing $K = 1000$ effective arrivals. Let $\hat{\lambda}_1^{(\ell)}$ be the estimate of $\lambda_1$ in the $\ell$-th run, and define
\[
\bar{\lambda}_1^{(L)}:= \frac{\sum_{\ell=1}^L \hat{\lambda}_1^{(\ell)}}{L} \qquad \sigma[\bar{\lambda}_1^{(L)}] := \sqrt{\frac{\sum_{\ell=1}^L (\hat{\lambda}_1^{(\ell)} - \bar{\lambda}_1^{(L)})^2}{L-1}},
\]
as the mean and standard deviation pertaining to the
estimates $\lambda_1$ resulting from $L$ runs. This applies to other unknown parameters. We represent our numerical results as $\left(\bar{\lambda}_1^{(L)}, \, \sigma[\bar{\lambda}_1^{(L)}]\right)$ in Table \ref{table:1}. For deeper analysis, we also present the average joining and switching rate, calculated as the mean of 
\[
\frac{K}{(\lambda_1+\lambda_2)T} \qquad \mbox{and} \qquad \frac{N}{K}\,
\]
from $L$ runs, where $T$ is the simulation time until we observe $K$ effective arrivals in each run, and $N$ is the number of switches in the $K$ joining customers in each run.

\begin{table}[h!] \caption{Estimates of $({\lambda}_1, {\lambda}_2, {\theta}, {c})$ from $1000$ runs.  The service times used are exponentially distributed with $\beta = 1, \, 5$, and Pareto distributed with $\beta = 2, \, 6$, for stations 1 and 2, respectively.} \label{table:1}
{\scriptsize
\begin{tabular}{c c c c c c c}
\toprule
True parameter & \multicolumn{4}{c}{Estimates} & average & average \\
$(\lambda_1, \lambda_2, \theta, c)$   & $\hat{\lambda}_1$ & $\hat{\lambda}_2$ & $\hat{\theta}$ & $\hat{c}$ & joining rate & switching rate
 \\ \hline
   \multicolumn{7}{c}{}\\
   \multicolumn{7}{c}{$G_1(x) = 1-e^{-x}, \, G_2(x) = 1-e^{-5x}$} \\ 
   \hline
   (1, 1, 1, 0.5) & (1.021, 0.084) & (1.020, 0.072) & (1.148, 0.296) & (0.489,  0.011) & 90.27\% & 12.99\%\\
   (1, 1, 3, 0.5) & (0.996, 0.079) & (0.981, 0.062) & (2.462, 0.387) & (0.479, 0.019) & 80.06\% & 6.10\%\\
   (1, 3, 1, 0.5) & (1.001, 0.116) & (3.006, 0.196)& (0.971, 0.456) & (0.487, 0.012) & 90.97\% & 7.19\%\\
   (1, 3, 3, 0.5) & (1.001, 0.113) & (2.992, 0.179)& (2.842, 0.557) & (0.475, 0.025) & 80.16\% & 3.25\%\\
   (1, 5, 1, 0.5) & (1.016, 0.153) & (4.974, 0.040)& (1.000, 0.320) & (0.489, 0.012) & 89.82\% & 5.68\%\\
   (1, 5, 3, 0.5) & (1.010, 0.132) & (4.987, 0.022)& (3.036, 0.394) & (0.478, 0.025) & 76.45\% & 2.64\%\\
    (5, 1, 1, 0.5) & (4.866, 0.184) & (0.996, 0.165) & (0.949, 0.184) & (0.498, 0.004) & 73.04\% & 49.82\%\\
    (5, 1, 3, 0.5) & (4.867, 0.182) & (0.989, 0.124) & (2.904, 0.233) & (0.496, 0.006) & 47.68\% & 30.15\%\\
   \hline
   \multicolumn{7}{c}{}\\
   \multicolumn{7}{c}{$G_1(x) = 1-\frac{1}{(1+x)^2}, \, G_2(x) = 1-\frac{1}{(1+x)^{6}}$} \\ 
      \hline
   (1, 1, 1, 0.5) & (1.019, 0.086) & (1.016, 0.079) & (1.105, 0.307) & (0.488, 0.012) & 89.07\% & 14.52\%\\
   (1, 1, 3, 0.5) & (0.992, 0.080) & (0.981, 0.064) & (2.614, 0.432) & (0.472, 0.027) & 78.93\% & 6.71\%\\
   (1, 3, 1, 0.5) & (1.001, 0.114) & (3.004, 0.205) & (1.005, 0.271) & (0.491, 0.010) & 85.32\% & 9.24\%\\
   (1, 3, 3, 0.5) & (0.999, 0.108) & (2.991, 0.189) & (2.970, 0.447) & (0.479, 0.022) & 75.54\% & 4.05\%\\
   (1, 5, 1, 0.5) & (1.003, 0.145) & (4.952, 0.068) & (1.006, 0.155) & (0.496, 0.007) & 77.24\% & 9.57\%\\
   (1, 5, 3, 0.5) & (1.010, 0.129) & (4.978, 0.033) & (2.979, 0.322) & (0.488, 0.017) & 66.85\% & 4.13\%\\
   (5, 1, 1, 0.5) & (4.912, 0.139) & (1.007, 0.192) & (0.979, 0.129) & (0.498, 0.004) & 66.38\% & 54.12\%\\
   (5, 1, 3, 0.5) & (4.923, 0.118) & (0.989, 0.124) & (2.958, 0.223) & (0.495, 0.006) & 45.03\% & 32.79\% \\
\bottomrule
\end{tabular}
}
\end{table}

\begin{figure}[h!]
    \centering
    \subcaptionbox{$\bar{\lambda}_1^{(L)} = 4.880 \, (\sigma[\bar{\lambda}_1^{(L)}] = 0.161)$}
    {\includegraphics[width=0.49\linewidth]{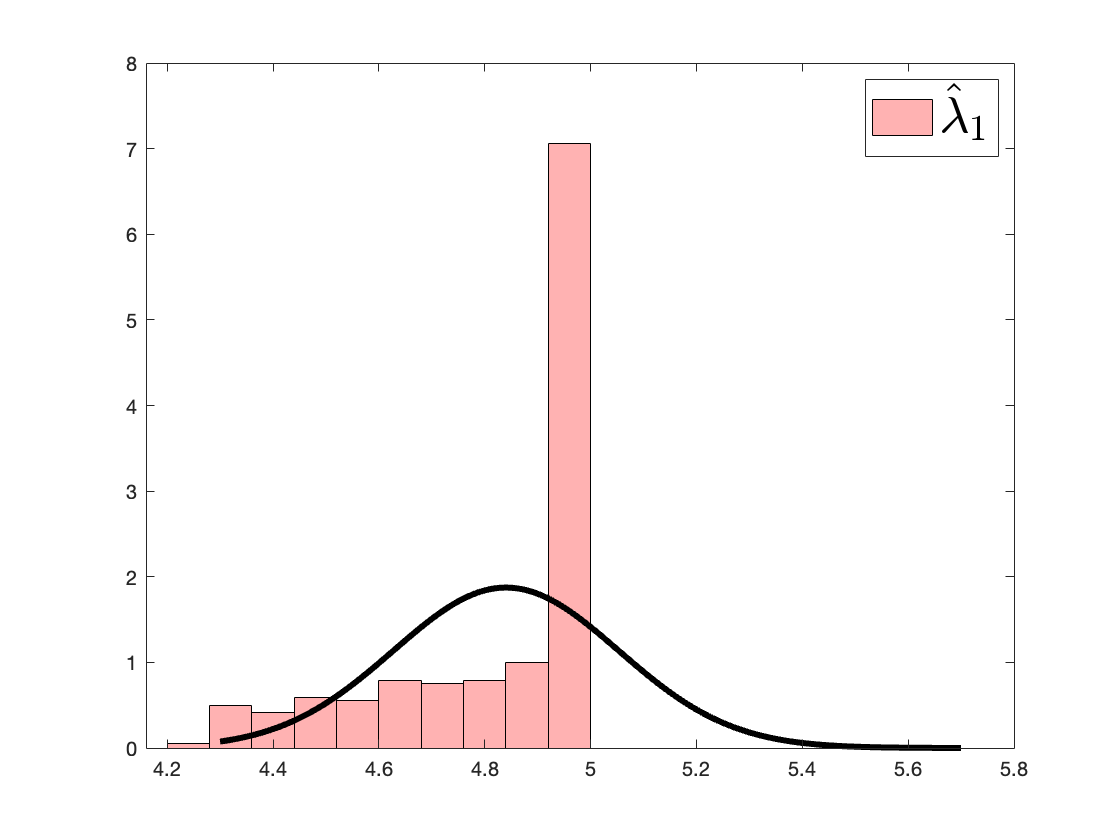}}
    \hfill
    \subcaptionbox{$\bar{\lambda}_2^{(L)} = 0.993 \, (\sigma[\bar{\lambda}_2^{(L)}] = 0.121)$}
    {\includegraphics[width=0.49\linewidth]{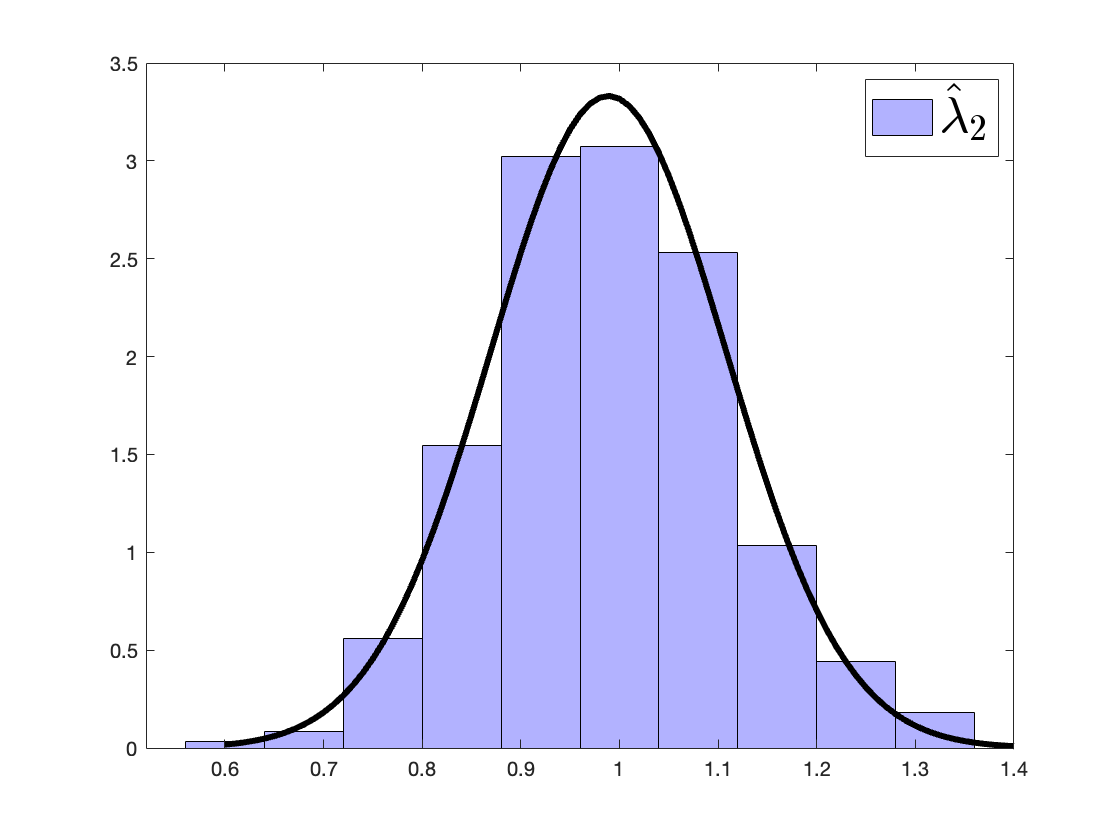}}
    \subcaptionbox{$\bar{\theta}^{(L)} = 2.939 \, (\sigma[\bar{\theta}^{(L)}] = 0.238)$}
    {\includegraphics[width=0.49\linewidth]{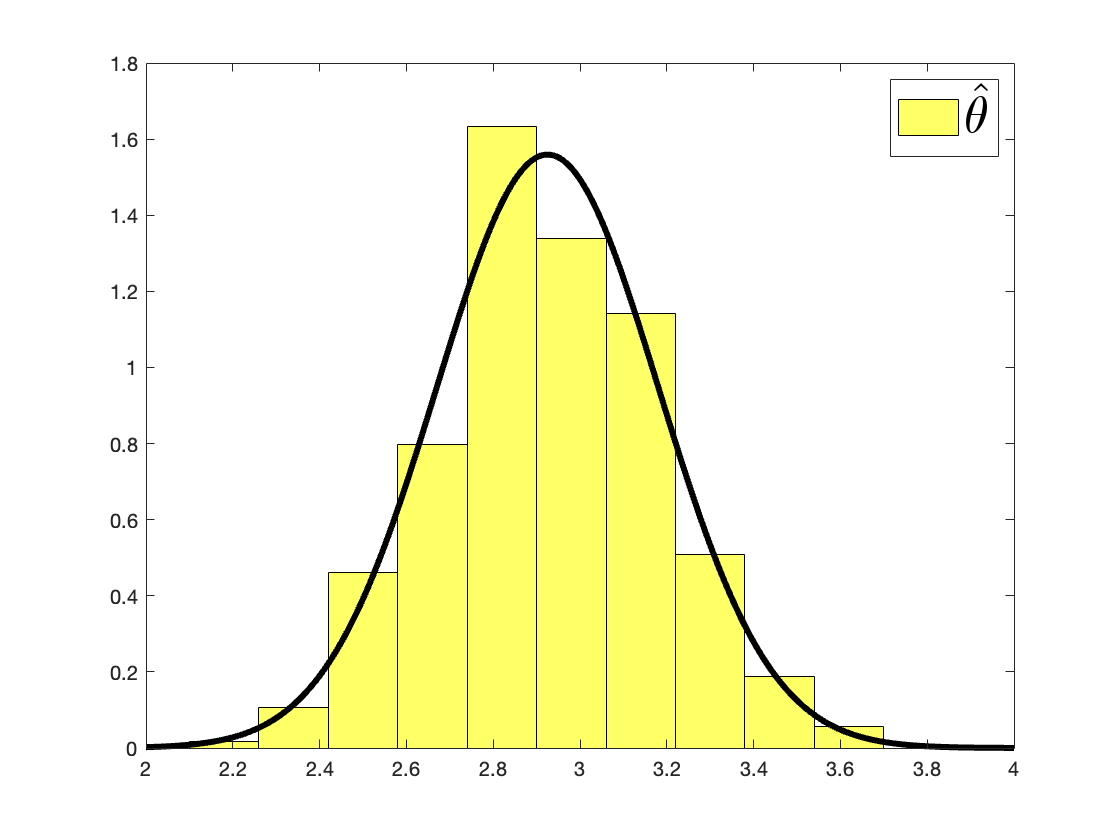}}
    \hfill
    \subcaptionbox{$\bar{c}^{(L)} = 0.498 \, (\sigma[\bar{c}^{(L)}] = 0.006)$}
    {\includegraphics[width=0.49\linewidth]{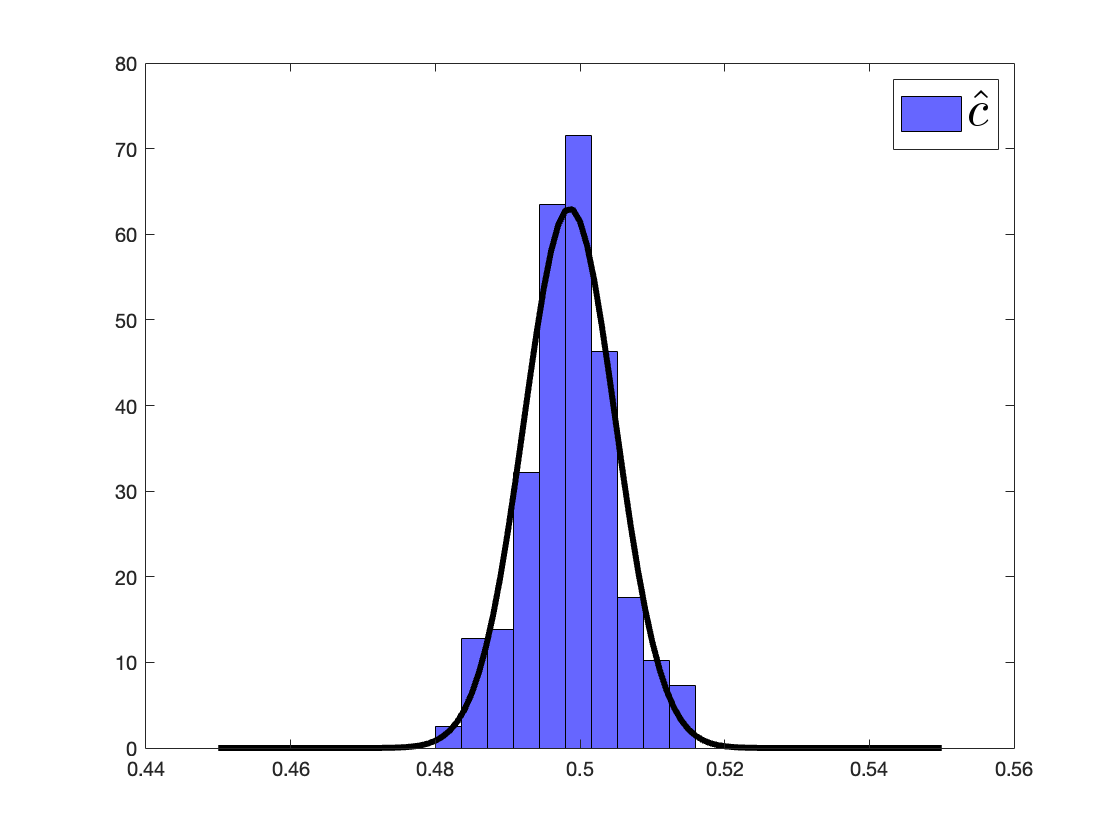}}
    \caption{Estimates from $L = 1000$ runs, given that the true value of $(\lambda_1, \lambda_2, \theta, c) = (5,1,3,0.5)$, and the service time distribution for server 1 and server 2 has $G_1(x) = 1-e^{-x}$, and $G_2(x) = 1-e^{-5x}$. The average joining and switching rates are 47.69\% and 29.94\%, respectively.}
    \label{fig:his1}
\end{figure}

\begin{figure}[h!]
    \centering
    \subcaptionbox{$\bar{\lambda}_1^{(L)} = 0.992 \, (\sigma[\bar{\lambda}_1^{(L)}] = 0.080)$}
    {\includegraphics[width=0.49\linewidth]{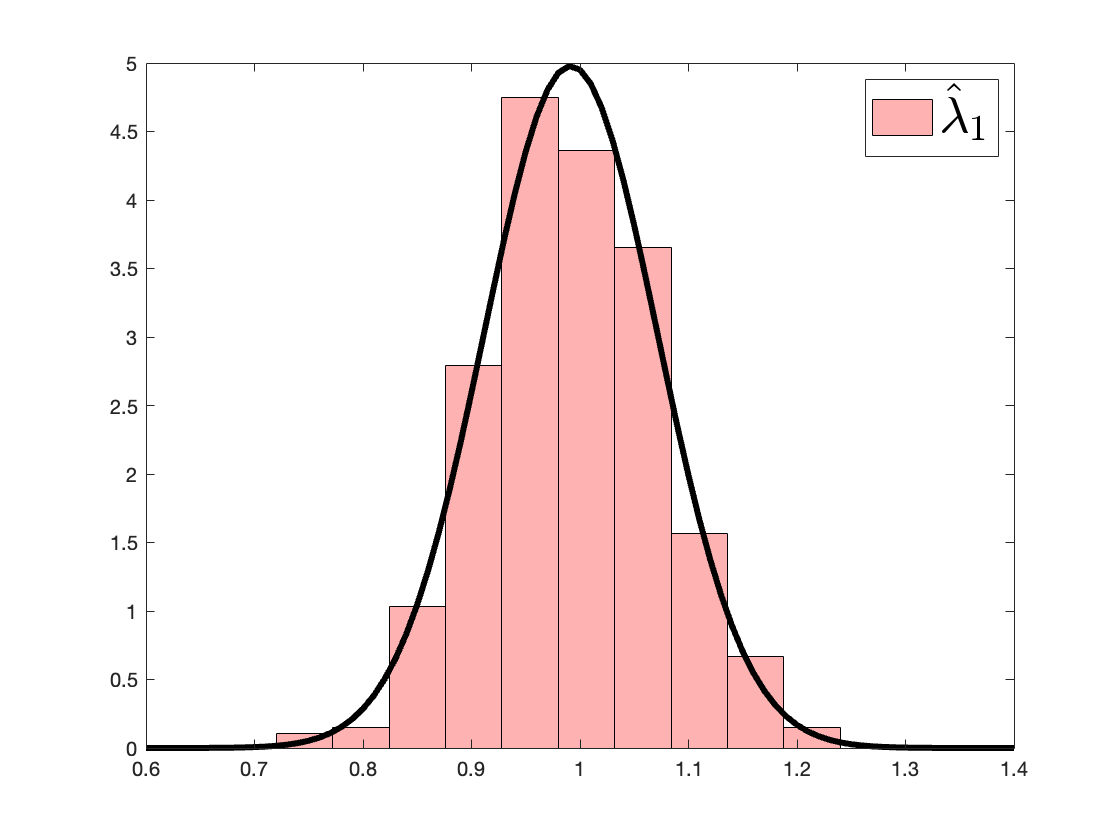}}
    \hfill
    \subcaptionbox{$\bar{\lambda}_2^{(L)} = 0.981 \, (\sigma[\bar{\lambda}_2^{(L)}] = 0.064)$}
    {\includegraphics[width=0.49\linewidth]{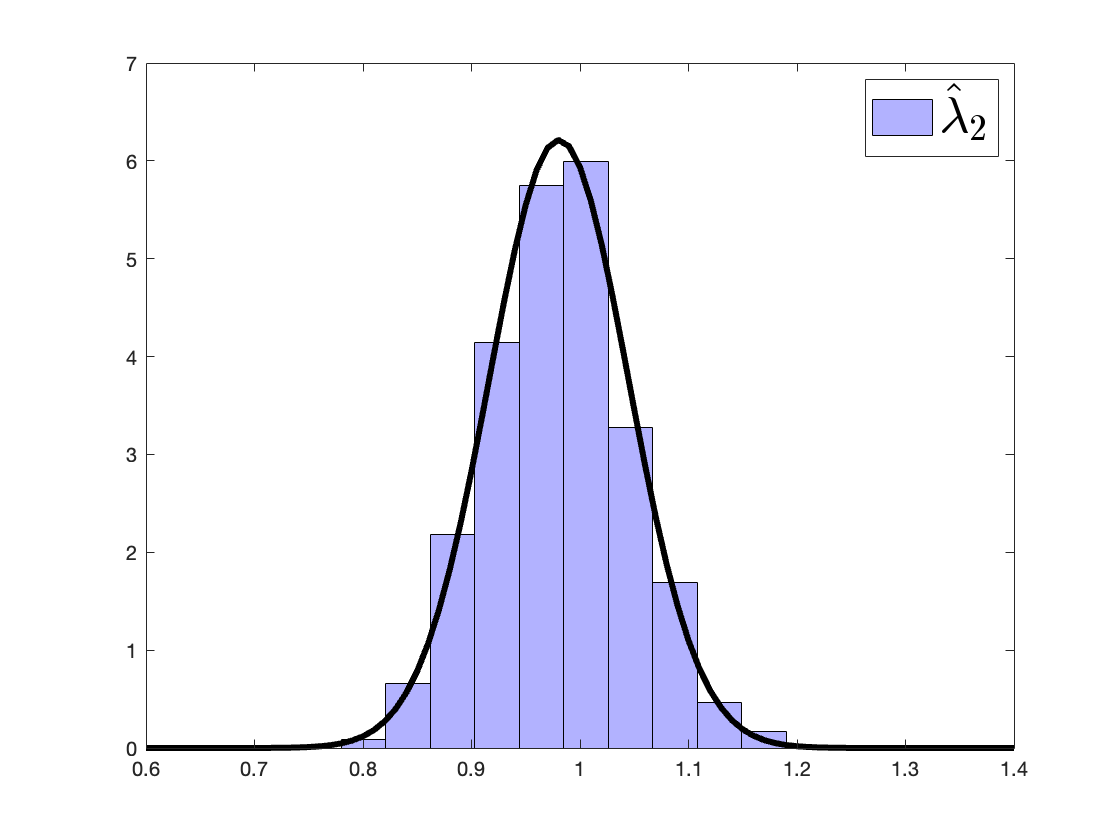}}
    \subcaptionbox{$\bar{\theta}^{(L)} = 2.614 \, (\sigma[\bar{\theta}^{(L)}] = 0.432)$}
    {\includegraphics[width=0.49\linewidth]{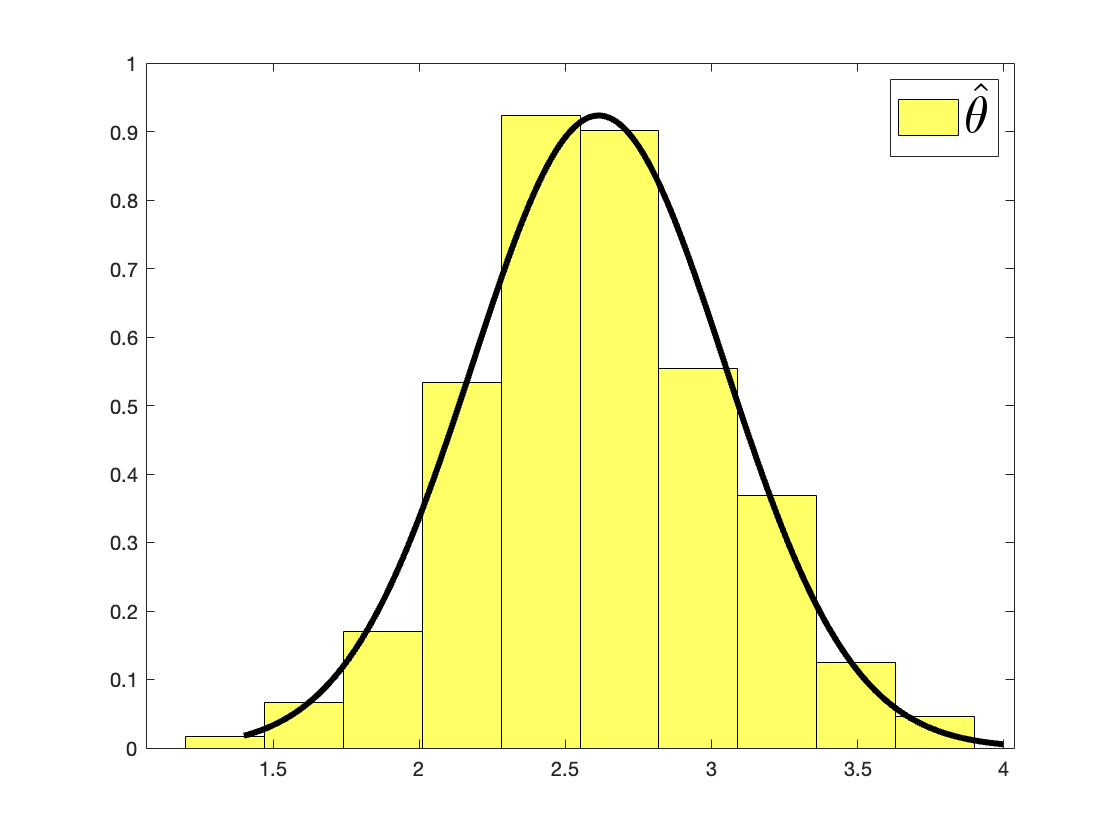}}
    \hfill
    \subcaptionbox{$\bar{c}^{(L)} = 0.472 \, (\sigma[\bar{c}^{(L)}] = 0.027)$}
    {\includegraphics[width=0.49\linewidth]{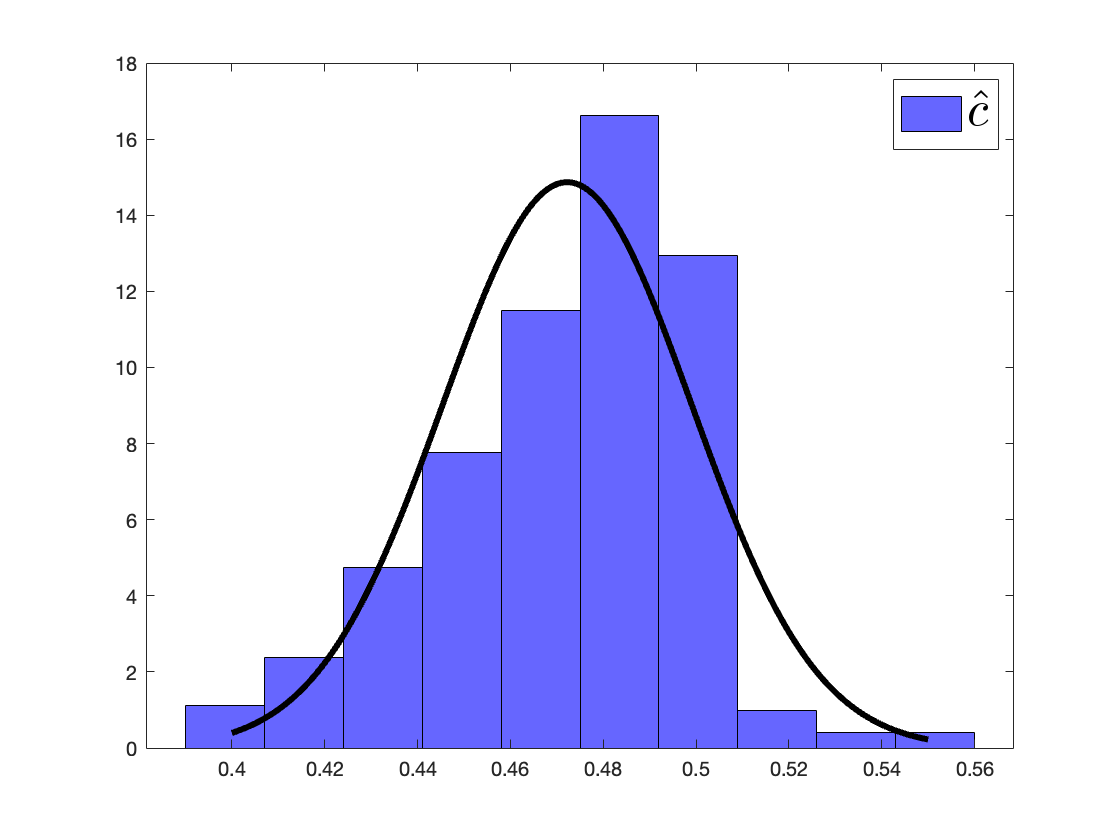}}
    \caption{Estimates from $L = 1000$ runs, given that the true value of $(\lambda_1, \lambda_2, \theta, c) = (1,1,3,0.5)$, and the service time distribution for server 1 and server 2 has $G_1(x) = 1-\frac{1}{(x+1)^2}$, and $G_2(x) = 1-\frac{1}{(x+1)^6}$. The average joining and switching rates are 78.93\% and 6.71\%, respectively.}
    \label{fig:his2}
\end{figure}

\begin{figure}[h!]
    \centering
    \subcaptionbox{$\bar{\lambda}_1^{(L)} = 1.018 \, (\sigma[\bar{\lambda}_1^{(L)}] = 0.138)$}
    {\includegraphics[width=0.49\linewidth]{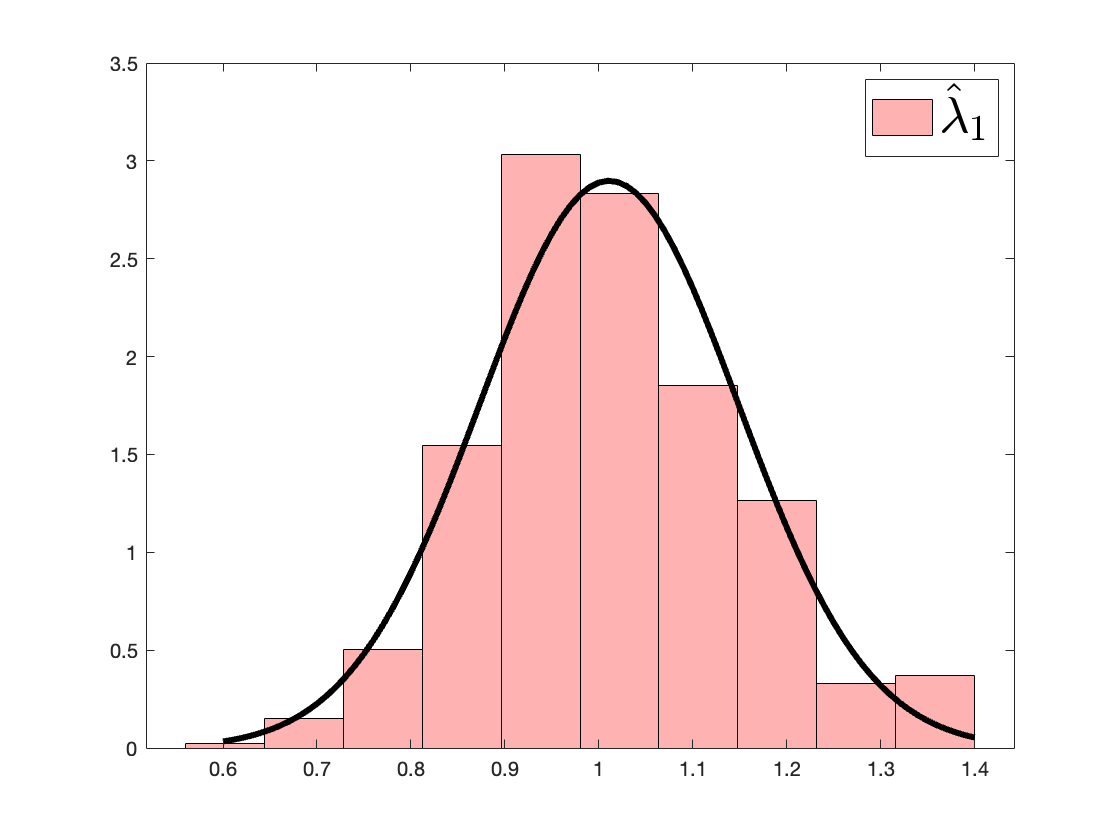}}
    \hfill
    \subcaptionbox{$\bar{\lambda}_2^{(L)} = 4.964 \, (\sigma[\bar{\lambda}_2^{(L)}] = 0.0.053)$}
    {\includegraphics[width=0.49\linewidth]{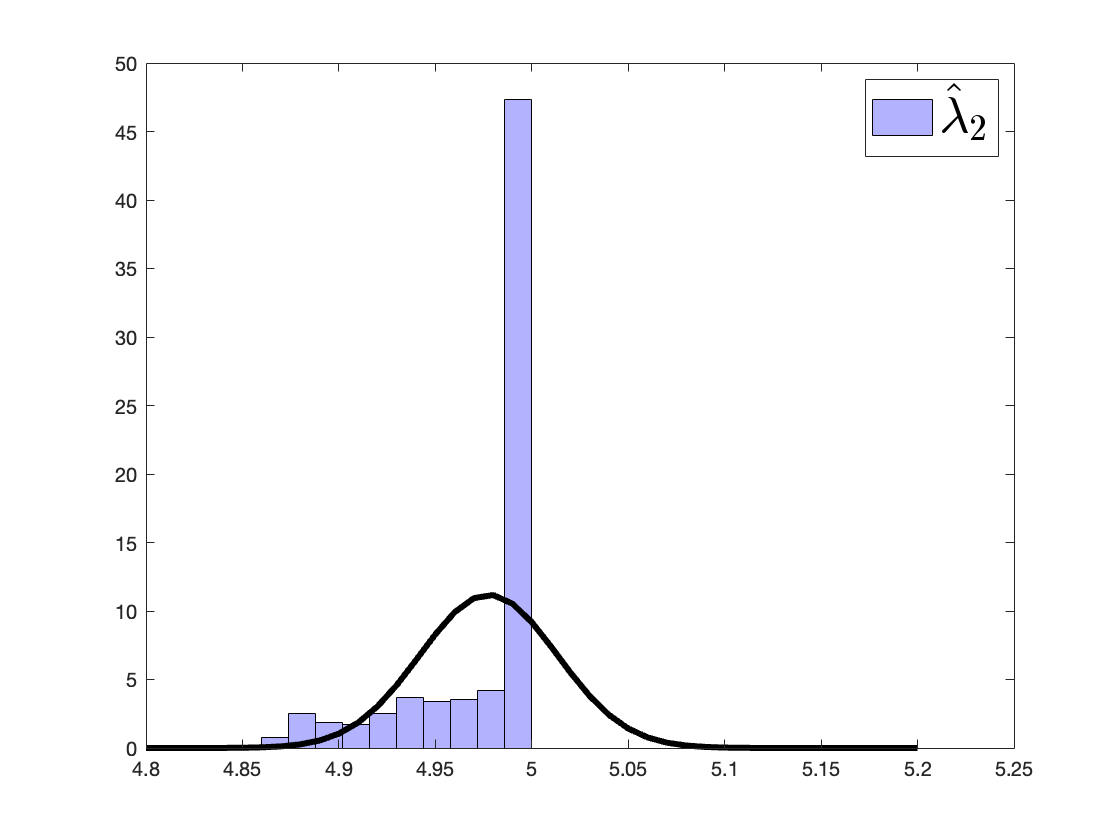}}
    \subcaptionbox{$\bar{\theta}^{(L)} = 3.004 \, (\sigma[\bar{\theta}^{(L)}] = 0.307)$}
    {\includegraphics[width=0.49\linewidth]{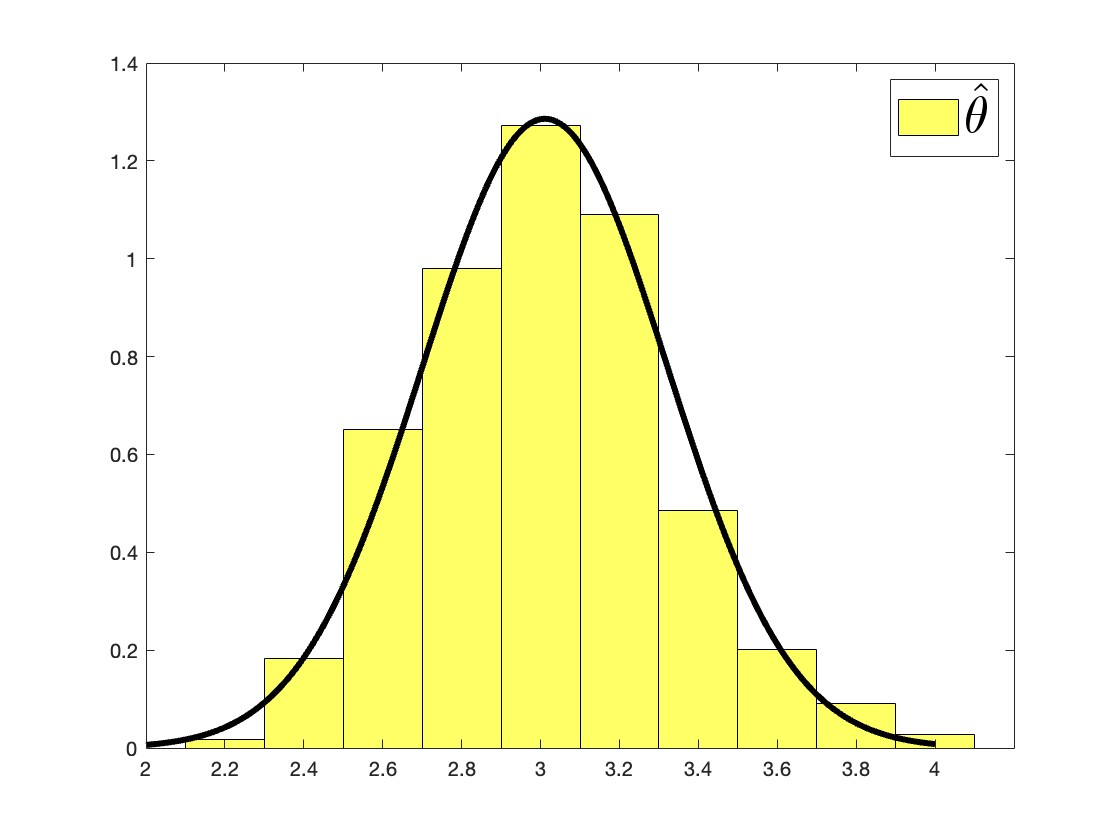}}
    \hfill
    \subcaptionbox{$\bar{c}^{(L)} = 0.488 \, (\sigma[\bar{c}^{(L)}] = 0.017)$}
    {\includegraphics[width=0.49\linewidth]{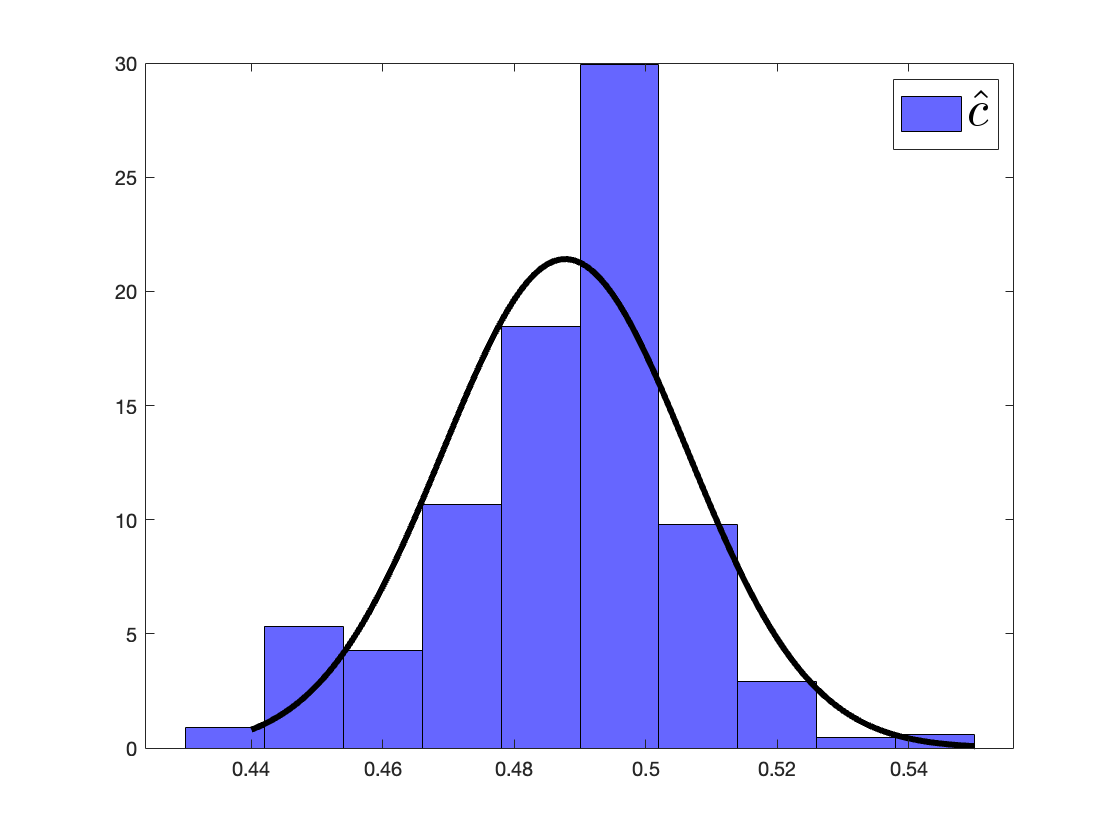}}
    \caption{Estimates from $L = 1000$ runs, given that the true value of $(\lambda_1, \lambda_2, \theta, c) = (1,5,3,0.5)$, and the service time distribution for server 1 and server 2 has $G_1(x) = 1-\frac{1}{(x+1)^2}$, and $G_2(x) = 1-\frac{1}{(x+1)^6}$. The average joining and switching rates are 67.07\% and 4.06\%, respectively.}
    \label{fig:his3}
\end{figure}

 Moreover, we plot the histograms of estimates of three cases, in Figure \ref{fig:his1}, \ref{fig:his2}, and \ref{fig:his3}, for comparison later. To check the asymptotic normality, we include a black line representing a normal distribution with the empirical mean and empirical standard deviation in the three figures. The code to generate the table and plots is in
{\small \url{https://github.com/encwang/TwoStationEstimation}}.

We have the following observations.
\begin{itemize}
    \item The accuracy of $\hat{c}$ improves with the number of observed switches, as more switches provide additional information about the switching cost. For example, in Figure \ref{fig:his2} and \ref{fig:his3}, estimates of $c$ are skewed towards smaller values, while in Figure \ref{fig:his1}, they are much closer to the true value. In Figure \ref{fig:his2} and \ref{fig:his3}, where switching rates are below 10\%, a higher estimate of $c$ value does not significantly reduce switching behavior, leading most estimates to fall below the true value.
    \item When the arrival rate is high, the system becomes very crowded, so further increases in the estimate of arrival rate have little effect on customer behavior. Although estimates of the higher arrival rate are relatively accurate, they tend to be skewed toward smaller values, as seen in estimates of $\lambda_1$ in Figure \ref{fig:his1} and $\lambda_2$ in Figure \ref{fig:his3}. 
    \item Conversely, when the arrival rate is low, servers are more likely to be idle, meaning that many arriving customers will join without factoring in the reward distribution. In other words, these joining observations provide limited information about the reward distribution, resulting in less accurate estimates of $\theta$, as shown in Figure \ref{fig:his2}.
\end{itemize}

Estimating $\lambda_1$ and $\lambda_2$ becomes more challenging when the average switching cost is low. To illustrate, consider a scenario where $c=0$, allows customers to switch freely. This implies that customer decisions are independent of the station they initially arrive at, making it impossible to infer the difference between $\lambda_1$ and $\lambda_2$, based on their joining behavior. 

There is a way to produce a lower bound on the switching cost $c$ via
\begin{equation} 
\tilde{c} := \max_{0 \le k \le K}\{ W^{(i_k)}_{k} -  W^{(-i_k)}_{k} \mid W^{(i_k)}_{k} > W^{(-i_k)}_{k}\}.
\end{equation}
If we observe a customer joining station $s$ while the virtual waiting time at station $s$ is higher than at station $-s$, it implies that the switching cost must exceed the waiting time difference; otherwise, the customer would have chosen to switch."

\section{Conclusion} \label{sec:conclusion}
In this paper, we examine the problem of estimating customers' perceived service value and potential demand in queueing systems where customers can choose to join, or switch between stations after joining. The challenge lies in two key aspects:
\begin{itemize}
\item Customers' perception of service value is private, making it challenging to identify the number of balking customers and, consequently, to estimate potential demand;
\item The switching cost is unknown, so the manager cannot determine whether a joining customer is local.
\end{itemize}

We managed to estimate the patience parameter, switching cost, and arrival rate to each station using only the workload process data from both stations—information typically available to the manager. The performance of our estimator is illustrated through numerical examples where the perceived service value follows a Pareto distribution, with results demonstrating its effectiveness.

Future research could explore several extensions. If observations focus on the number of customers at each station rather than exact waiting times, an arriving customer's decision will hinge on expected waiting times, making the service time distribution a crucial factor. Furthermore, if switching costs are incurred gradually over time—similar to travel costs—then transitions occurring during the travel period must also be considered. A key challenge here is that these transitions are influenced by the concurrent decisions of others.

\section{Acknowledgment}
The authors would like to thank Opher Baron and Philipp Afech for their valuable comments and advice. This research was supported by the European Union’s Horizon 2020 research and innovation programme under the Marie Sklodowska-Curie grant agreement no.\ 945045, and by the NWO Gravitation project NETWORKS under grant agreement no.\ 024.002.003. \includegraphics[height=1em]{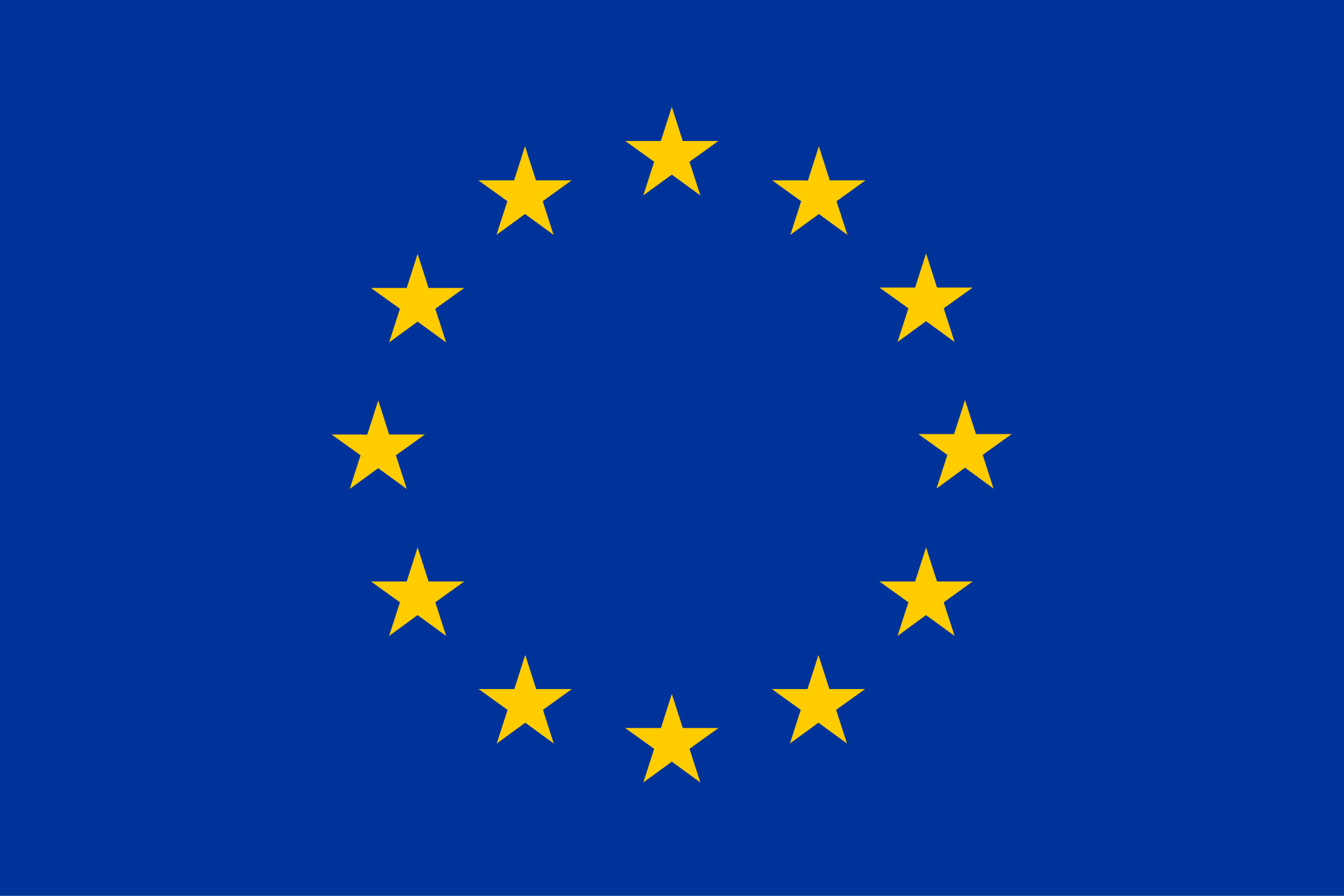}

\end{document}